\documentstyle[a4,12pt]{article}
 
\newcommand{\be}{\begin{equation}}
\newcommand{\bea}{\begin{eqnarray}}
\newcommand{\bean}{\begin{eqnarray*}}
\newcommand{\ee}{\end{equation}}
\newcommand{\eea}{\end{eqnarray}}
\newcommand{\eean}{\end{eqnarray*}}
\newcommand{\ZZ}[1]{{\bf Z}_{#1}}

\newcommand{\ol}{\overline}
\newcommand{\half}{\frac{1}{2}}
\newcommand{\Tr}{\mbox{Tr}}
\newcommand{\A}{{\cal A}}
\newcommand{\th}[2]{\vartheta_{#1}(#2|t)}

\newcommand{\bmat}{\left(\begin{array}}
\newcommand{\emat}{\end{array}\right)}

\newcommand{\e}[2]{e^{\frac{#1 \pi i}{#2}}}
\newcommand{\expm}[2]{e^{\pm\frac{2 \pi i #1}{#2}}}
\newcommand{\exmp}[2]{e^{\mp\frac{2 \pi i #1}{#2}}}
\newcommand{\eb}[2]{e^{-\frac{#1 \pi i}{#2}}}
\newcommand{\dia}{\mbox{diag}}
\newcommand{\id}[1]{{\bf I}_{#1}}

\newcommand{\fr}[2]{\frac{#1}{#2}}
\newcommand{\frn}{\frac{n}{N}}
\newcommand{\frm}{\frac{m}{M}}
 
\newcommand{\nn}{\nonumber}

\def\NPB#1#2#3{{\it Nucl.\ Phys.}\/ {\bf B#1} (19#2) #3}
\def\PLB#1#2#3{{\it Phys.\ Lett.}\/ {\bf B#1} (19#2) #3}
\def\PRD#1#2#3{{\it Phys.\ Rev.}\/ {\bf D#1} (19#2) #3}

\def\JGP#1#2#3{{\it Jour.\ Geom.\ Phys.}\/ {\bf #1} (19#2) #3}
\def\etal{{\it et al.\/}}

\begin{document}

\addtolength{\baselineskip}{.5mm}
\renewcommand{\theequation}{\thesection.\arabic{equation}}
\renewcommand{\thefootnote}{\fnsymbol{footnote}}

\begin{flushright}
August 1997\\
{\sc thu}-97/21\\
{\sc hep-th}/9708040\\
\end{flushright}
\vspace{2cm}
\thispagestyle{empty}
\begin{center}
{\large{Four-dimensional $N=1$ $\ZZ{N}\times\ZZ{M}$ Orientifolds 
}}\\[18mm]
{\sc Gysbert Zwart\footnote{e-mail address: zwart@fys.ruu.nl, telephone: ++31 
30 2533058, fax: ++31 30 2531137} }\\[7.5mm]
{\it Institute for Theoretical Physics,\\ University of Utrecht,\\ P.O. Box 
80006, 3508 TA  Utrecht,\\ The Netherlands}\\[18mm]
 
{\sc Abstract}
\end{center}
We calculate the tadpole equations and their solutions for
a class of four-dimensional orientifolds with orbifold group
$\ZZ{N}\times\ZZ{M}$,
and we present the massless bosonic spectra of these models. Surprisingly
we find no consistent solutions for the models with $\ZZ2\times\ZZ4$ and
$\ZZ4\times\ZZ4$ orbifold groups.
\\

\\
\vfill

\pagebreak

\section{Introduction}

In the search for string compactifications to four dimensions breaking
supersymmetry to $N=1$, orbifolds have played an important role, as they
yield solvable models \cite{DHVW}. In the past attention focused on
heterotic
orbifolds \cite{Ibetal}. With the advent of D-branes, (see \cite{Dbranes}
and references
therein), a different class of theories came under investigation, where
the orbifold group includes a world sheet symmetry: orientifolds
\cite{ori,GP}. These are models in which, beside unoriented closed
strings, open strings ending on various types of D-branes occur.

The simplest example of such a model is the type I string theory. This theory
can be viewed as an orientifold of the type IIB string theory. Type IIB 
strings are symmetric under an exchange of their left and right moving 
sectors, and this symmetry can be divided out. The first effect of this 
is a truncation of the closed string spectrum to those states invariant 
under the symmetry; this truncated theory, however, is inconsistent due 
to non-vanishing tadpoles. The situation is cured by adding open strings, 
ending on space-filling nine-branes, and supporting an $SO(32)$ gauge 
group. 

In \cite{PW} evidence was presented for a duality relating this type I 
theory to heterotic strings with the same gauge group. Studying type I 
orbifolds therefore provides a means for investigating non-perturbative 
aspects of the heterotic orbifolds considered previously. So far most 
attention has been paid to six-dimensional compactifications, where 
orientifolds typically describe a phase involving more than one tensor 
multiplet, a situation that is out of reach of perturbative heterotic 
theory. 

In these lower dimensional orientifold models, obtained by dividing 
out symmetries from toroidally compactified type I
theories, apart from nine-branes also five-branes have to be introduced.
It was found by Gimon and Polchinski \cite{GP} that in 
these models further consistency conditions have to be met, due to the 
interplay between open strings ending on nine-branes and five-branes. Their 
work on the $\ZZ2$ $K3$ orientifold was subsequently extended
to other $K3$ orientifolds in
\cite{GJ} and \cite{DP}, and to a four-dimensional model with orbifold
group $\ZZ2\times\ZZ2$ in \cite{BL}. In the latter case three differently
oriented sets of five-branes appear, with more stringent restrictions on
the representation of the group elements in the open string sectors,
arising from the mixing between different sectors. Further
four-dimensional orientifolds were presented in \cite{K,KS}.

Here we extend the collection of four-dimensional orientifolds with $N=1$
supersymmetry to models with spacetime orbifold groups of the form
$\ZZ{N}\times\ZZ{M}$. Apart from the spacetime orbifold group elements, 
which are of the form $\alpha^n\beta^m$, with $\alpha$ ($\beta$) the
generator of $\ZZ{N}$ ($\ZZ{M}$), the orientifold group involves the
elements $\Omega\alpha^n\beta^m$, where $\Omega$ is the world sheet
orientation reversal. This introduces orientifold fixed planes,
necessitating the addition of D-branes to compensate their RR-charges. 
We find that the algebraic consistency conditions that have to be imposed 
are, in some cases, too restrictive to have a solution. 

The plan of the paper 
is as follows. We first recall the general construction
of orientifold models. Then we give representations of the action of the
orientifold group
elements on the open string Chan-Paton factors, consistent with tadpole
cancellation and further algebraic consistency conditions of the type
presented first in \cite{GP}. This generalises similar solutions in
\cite{GJ,DP,BL}. In the next subsection the closed string untwisted 
and twisted states are calculated. Finally we use the solutions of the
Chan-Paton representations to obtain the open string spectra. The results
of the tadpole computations are presented in an appendix. 

\section{Four-dimensional $\ZZ{N}\times\ZZ{M}$-orientifolds}

\subsection{Orientifolds}

Let us briefly recall the general features of orientifolds \cite{GP}. One
starts 
with a type IIB compactification on an orbifold. This model has a 
symmetry $\Omega$, exchanging left and right movers on the world sheet. 
Dividing out this symmetry produces the orientifold. The closed string 
spectrum consists of those states invariant under $\Omega$. In addition 
one has to include open strings ending on Dirichlet branes; these 
D-branes are necessary to cancel the charge of the orientifold 
planes, the fixed planes under the orientifold group elements. 

The number of D-branes, as well as the action of the group elements on 
them, can be determined by a calculation of the amplitudes for the 
exchange of massless antisymmetric tensor particles from the RR-sector, 
which couple to the charges of orientifold planes and D-branes. The total 
contribution for each species of these particles should vanish in the 
exchange between the boundary states consisting of both orientifold 
planes and D-branes, since this amplitude is proportional to the square 
of the total charge.

The action of the group elements on the D-branes is encoded in the 
transformation of the Chan-Paton matrices $\lambda_{ij}$ associated to
open strings 
ending on D-branes. Under a symmetry $\alpha$, these Chan-Paton matrices 
are conjugated by a unitary matrix $\gamma_\alpha$,
\be
\lambda\to \gamma_\alpha\lambda\gamma_\alpha^{-1}
\ee 
These $\gamma_\alpha$ should form a projective representation of the group. 
The cancellation of the charges, or tadpole conditions, places 
restrictions on the traces of (products of) the $\gamma_\alpha$. 
Furthermore there are restrictions originating from the fact that 
quantities associated to symmetries are conserved in string interactions. 
The prototype of these constraints is the fact that in the presence of 
nine-branes with $\gamma_\Omega$ symmetric, five-branes have 
antisymmetric $\gamma_\Omega$; this is a consequence of the fact that 
both sectors are related via $59$-strings \cite{GP}.

One then determines the structure of the Chan-Paton matrices invariant 
under the action of the symmetry group. This yields the gauge 
group carried by the open string vectors and the representations of 
the open string matter.

\subsection{The orientifold groups}

The models we calculate are orientifolds 
with four-dimensional $N=1$ 
supersymmetry. They arise from type IIB string theory compactified on a 
six-dimensional torus modded out by a symmetry group 
$\ZZ{N}\times\ZZ{M}$, as well as the orientation reversal symmetry 
$\Omega$. The group is abelian and its elements are therefore of the form 
$\Omega\alpha^n\beta^m, \alpha^n\beta^m$, with $\alpha,\beta$ generators 
of the two cyclic groups.

Parametrising the torus by the three complex numbers $z_1,z_2$ and $z_3$, 
we choose the action of the groups $\ZZ{N}$ and $\ZZ{M}$ as 
\be
(z_1,z_2,z_3)\to (e^{\frac{2\pi i n}{N}}z_1, e^{-\frac{2\pi
i n}{N}}z_2,z_3),\mbox{ resp. } (z_1,z_2,z_3)\to (z_1,e^{\frac{2\pi i
m}{M}}z_2, e^{-\frac{2\pi i
m}{M}}z_3).
\ee
For this to be a symmetry, we choose the three two-tori to be either the 
$SU(2)\times SU(2)$ or the $SU(3)$ root lattice. These are then invariant 
under the groups $\ZZ2,\ZZ4$ resp. $\ZZ2,\ZZ3,\ZZ6$. (The model with
$N=M=3$ was already considered in \cite{KS}.) The action on the 
spinors is defined by $e^{\frac{2\pi i
n}{N}(J_{45}-J_{67})}$, $e^{\frac{2\pi i m}{M}(J_{67}-J_{89})}$. From 
this we
can determine that the unbroken supercharges are given by  spinors 
$(s_0,s_1,s_2,s_3,s_4)$, with $s_3=s_4=s_5$. Since they are also
GSO-projected (chiral) this leaves us with four components, or one spinor
in four dimensions. Due to the $\Omega$ projection left and rightmovers
are identified, so we have $N=1$ in $d=4$.

Since in every model $\Omega$ is one of the group elements, they will all 
contain nine-branes. Five-branes also occur for those models with 
additional elements of order two. 
 
\subsection{Solutions to the tadpole equations}
\label{sec:solns}

The tadpole equations are obtained from the calculation of closed string 
massless RR-sector exchange between boundary states consisting of all 
orientifold planes and D-branes. This calculation is most easily done by 
viewing the closed string tree-level exchange diagrams as open string one 
loop diagrams. The different boundary conditions for orientifold planes 
and D-branes give rise to cylinder, M\"obius strip and Klein bottle 
diagrams, from which the part associated to the required exchange can be 
easily extracted. The details and results of these calculations are 
deferred to the appendix.

Here we turn to the solutions of the tadpole equations. These can be 
obtained by generalising those found in \cite{GJ,BL}. We will start with 
the simplest case of $\ZZ3\times\ZZ3$, where only nine-branes are 
present. The trace of the unit matrix is $32$, as will be the case in all 
models. The number of dynamical nine-branes, however, is smaller, as
these $32$ 
include the various images under the symmetries. The form of the tadpoles
in this case is given in eqs. (\ref{eq:oo1},\ref{eq:oo2}). The equations
are
solved by
gamma-matrices 
\be
\gamma_{\fr13,0}=\dia (\e23,\eb23,1,1)\otimes\id8,\;
\gamma_{0,\fr13}=\dia (\eb23,1,\e23,1,\eb23,1,\e23,1)\otimes\id4,
\ee
(where the indices indicate the elements of the two orbifold groups)
together with the representation
\be
\gamma_\Omega=\bmat{cccc}0&1&0&0\\1&0&0&0\\0&0&0&1\\0&0&1&0\emat\otimes\id8,
\ee
which swaps conjugate numbers in both matrices. The general $n,m$
matrices are simply obtained by defining
$$
\gamma_{n,m}=\gamma_{\fr13,0}^n\gamma_{0,\fr13}^m.
$$
 
The next simplest models are $\ZZ3\times\ZZ6$ and $\ZZ2\times\ZZ3$. In
this case we have, in 
addition to
the nine-branes, $32$ five-branes stretched along the $0\ldots 5$
dimensions. The tadpoles that should vanish in this case are given in
eqs. (\ref{eq:eo1}--\ref{eq:eo4}). For the
nine-branes we find 
\be
\gamma_{0,\fr16,9}=\dia (i,-i)\otimes\dia (\eb23,\e23,1,1)\otimes\id4,\;
\gamma_{\fr13,0,9}=\id4 \otimes\dia (\eb23,1,\e23,1)\otimes\id2,
\ee
so that the $\gamma_{0,\fr13,9}$ and $\gamma_{\fr13,0,9}$ are $-1$ and $+1$ 
times the matrices of the previous case. For $\gamma_{0,\fr16,5}$ we can take
$\gamma_{0,\fr16,9}$, and $\gamma_{\fr13,0,5}=-\gamma_{\fr13,0,9}$. 
Further we can again take the matrix $\gamma_{\Omega,9}$ to exchange all  
conjugate entries,
i.e.
\be
\gamma_{\Omega,9}=\bmat{cc}0&1\\1&0\emat\otimes\bmat{cccc}0&1&0&0\\
1&0&0&0\\0&0&0&1\\0&0&1&0\emat\otimes\id4.
\ee  
For $\gamma_{\Omega,5}$ we can take $\gamma_{\Omega,9}\gamma_{0,\fr12,9}$. 
This is
antisymmetric, as it should be.

For the $\ZZ2\times\ZZ3$ model's solution we can use the same solution, 
discarding those elements not belonging to the $\ZZ2$ subgroup of $\ZZ6$. 

We now arrive at the somewhat more complicated models involving three 
different sets of five-branes: $32$ $5_1$-branes along the $4-5$ plane, $32$ 
$5_2$-branes along 
the $6-7$ plane, and $32$ $5_3$-branes along the $8-9$ plane.  The basic 
example of 
this kind, with orbifold group $\ZZ2\times\ZZ2$, was explored in \cite{BL}. 
The algebraic consistency conditions relating the Chan-Paton 
representation of the orientifold group are more involved,  
creating $\gamma$'s in the different sectors (nine-branes and different 
five-branes) that do not commute. Here we will employ the solution of 
\cite{BL} to construct the generalisation to the models under consideration.

The tadpole equations are obtained by setting
the expressions
(\ref{eq:ee1}--\ref{eq:ee6})
to zero. For the $\ZZ6\times\ZZ6$ model, we find for the nine-branes
\bea
\gamma_{0,\fr16,9}&=&i\sigma_2\otimes\id2\otimes\dia(\e23,\eb23,1,1) ,\\
\nn
\gamma_{\fr16,\fr16,9}&=&-\sigma_3\otimes 
i\sigma_2\otimes\dia(\e23,1,\eb23,1) 
,\\ \nn  \gamma_{\fr16,0,9}&=& \gamma_{1,1,9}\gamma_{0,1,9}^{-1}.
\eea
with the appropriate identity factors. This choice gives the 
answers of \cite{BL}
when the third power is taken. The five-brane matrices can be constructed
similarly:
\bea
\gamma_{0,\fr16,5_1}&=&i\sigma_2\otimes\id2\otimes\dia(\e23,\eb23,1,1) ,\\
\nn
\gamma_{\fr16,\fr16,5_1}&=&\sigma_3\otimes\id2\otimes\dia(\e23,1,\eb23,1) 
,\\ \nn
\gamma_{\fr16,0,5_1}&=& \gamma_{\fr16,\fr16,5_1}\gamma_{0,\fr16,5_1}^{-1}.
\eea
\bea
\gamma_{0,\fr16,5_2}&=&-\sigma_1\otimes\id2\otimes\dia(\e23,\eb23,1,1) ,\\
\nn
\gamma_{\fr16,\fr16,5_2}&=&i\sigma_2\otimes\id2\otimes\dia(\e23,1,\eb23,1) 
,\\ \nn
\gamma_{\fr16,0,5_2}&=& -\gamma_{\fr16,\fr16,5_1}\gamma_{0,\fr16,5_1}^{-1}.
\eea
\bea
\gamma_{0,\fr16,5_3}&=&\sigma_3\otimes\id2\otimes\dia(\e23,\eb23,1,1) ,\\
\nn
\gamma_{\fr16,\fr16,5_3}&=&-\sigma_1\otimes\id2\otimes\dia(\e23,1,\eb23,1) 
,\\ \nn
\gamma_{\fr16,0,5_3}&=& \gamma_{\fr16,\fr16,5_1}\gamma_{0,\fr16,5_1}^{-1},
\eea
while the representation for $\Omega$ is $\id{}$ times the 
$\ZZ3\times\ZZ3$ structure for 
nine-branes and $\id{}\otimes i\sigma_2$ times this structure for all the 
five-branes.
Again we can find the $\ZZ2\times\ZZ6$ (and the $\ZZ2\times\ZZ2$) model 
by restriction from these matrices.

For the models involving $\ZZ4$ on the other hand, it turns out we cannot 
satisfy the consistency conditions. Let us take $\ZZ4\times\ZZ4$ as an 
example, and concentrate on the gammas in the nine-brane sector. We may 
again choose $\gamma_\Omega$ to be the identity matrix. This then 
determines $\gamma_{\fr12,0},\gamma_{0,\fr12}$ and $\gamma_{\fr12,\fr12}$ 
to be antisymmetric matrices, as in the $\ZZ2\times\ZZ2$ model of
\cite{BL}. To see this, one uses that the tadpole conditions require
e.g. $\gamma_{\Omega,\fr12,0}$ to be symmetric in the $5_3$ sector. From
consideration of the process where two $95_3$ states go into a $99$ or a
$5_35_3$ state, one obtains that $\gamma_{\Omega,\fr12,0}$ should have the
opposite symmetry property on the nine-branes. Since we chose
$\gamma_\Omega=\id{}$ in this sector, we find that also
$\gamma_{\fr12,0}$ should be antisymmetric in the $9$ sector. All matrices 
are required to 
form a projective representation of the orbifold group, so that we have
\be
\label{eq:z4}
\gamma_{\fr14,0}^4=a \id{},\quad \gamma_{0,\fr14}^4=b\id{},\mbox{ 
and }c\gamma_{\fr14,0}\gamma_{0,\fr14}=  \gamma_{0,\fr14}\gamma_{\fr14,0}.
\ee
Here $a,b$ and $c$ are arbitrary phases. 
Since $\gamma_{\fr14,0}^2\gamma_{0,\fr14}^2$ should equal, up to a
phase, $\gamma_{\fr12,\fr12}$, it is antisymmetric. From this and the
third identity in (\ref{eq:z4}), we find 
that $c^4=-1$. But as 
$\gamma_{0,\fr14}\gamma_{\fr14,0}\gamma_{0,\fr14}^{-1}= 
c\gamma_{\fr14,0}$ (again the third equation in (\ref{eq:z4})), we also
have, by conjugating this three times more, that 
$\gamma_{0,\fr14}^4\gamma_{\fr14,0}\gamma_{0,\fr14}^{-4}= 
c^4\gamma_{\fr14,0}$. However, if we now use the second identity in
(\ref{eq:z4}), the phase $b$ drops out and we obtain that $c^4=1$.
The same contradiction can be 
derived for the $\ZZ2\times\ZZ4$ case. So we have found the surprising 
result that the consistency conditions do not admit a solution to these 
models, which at first sight seem to be as reasonable as the others.

This concludes the solutions to the tadpole equations.

\subsection{The massless closed string spectra}

The spectrum of the models consists of closed and open string states. The 
closed string states are those type IIB orbifold states that are 
invariant under $\Omega$. The states group together in multiplets of the 
$N=1$ supersymmetry. 

The closed string spectrum in the untwisted sector is built up from the
following massless states:
\\

\noindent 
\begin{tabular}{|r|l|c|l|}
\hline
Sector & state & $(z,w)$ & helicity\\ \hline
NS & $\psi^\mu|0>$ & $1$ & $\pm 1$ \\
   & $\psi^1|0>$ & $e^{\pm2\pi i z}$ & $2\times 0$\\
   & $\psi^2|0>$ & $e^{\pm 2\pi i (w-z)}$ & $2\times 0$\\
   & $\psi^3|0>$ & $e^{\pm 2\pi i w}$ & $2\times 0$\\
\hline
R  & $|s_1=s_2= s_3=s_4>$ & $1$ & $\pm\half$ \\
   & $|s_1=s_2=-s_3=-s_4> $& $ e^{\pm 2\pi i z}$ & $\pm\half $\\
   & $|s_1=-s_2= s_3=-s_4>$& $ e^{\pm 2\pi i (z-w)}$& $\pm\half$\\
   & $|s_1=-s_2=-s_3=s_4> $& $e^{\pm 2\pi i w}$& $\pm \half$ \\
\hline
\end{tabular}
\\

\noindent The three internal tori are labelled by the numbers $1,2,3$. The 
$(z,w)$ column lists the behaviour of the 
states under the element $z=\fr{n}{N}$ of $\ZZ{N}$ and $w=\fr{m}{M}$ of 
$\ZZ{M}$. In the final column the helicities of the states
are listed. Only states surviving the 
GSO projection are given. The bosonic closed string states are 
those combinations invariant under the orbifold group, and 
left-right symmetric in the NSNS sector, antisymmetric in the RR 
sector. For the $\ZZ3\times\ZZ3,\ZZ3\times\ZZ6$ and $ \ZZ6\times\ZZ6$ 
models this gives the massless bosonic spectrum  
\bean
NSNS:&& \pm 2 + 4\times 0\\
RR  :&& 4\times 0 ,
\eean
making up the gravity multiplet and four chiral multiplets; 
for $\ZZ2\times\ZZ3$ and $\ZZ2\times\ZZ6$ we find
\bean
NSNS:&& \pm 2 + 6\times 0\\
RR  :&& 4\times 0 ,
\eean
which gives five chiral multiplets.

Next we determine the massless twisted closed string sectors. To this end 
we determine the twisted contributions to the cohomology of the 
orbifold model, which give the twisted RR ground states of the IIB
orbifold 
compactification \cite{VW}. Of these states only half survive the 
orientifold projection $\Omega$, and these sit together with an equal 
number of NSNS states in chiral supermultiplets. We carry out explicitly 
only the calculation of the $\ZZ3\times\ZZ3$ and $\ZZ2\times\ZZ3$ models, 
and just give the results of the others.

For the orientifold with orbifold group $\ZZ3\times\ZZ3$ we have the 
group generators $\alpha$, acting on the complex torus coordinates with 
charges $(\fr13,\fr23,0)$, and $\beta$ : $(0,\fr13,\fr23)$. First we have 
states twisted by $\alpha$, whose fixed point set is 9 tori; of   
the torus cohomology only half is invariant under $\beta$; in total we end up
with one chiral multiplet for each such torus; there are 6 order three 
elements like
$\alpha$, so this gives $54$ multiplets. Then we have states twisted by
$\alpha^2\beta$, which have $27$ fixed points; there are two such twisted 
sectors,
giving in total $27$ chiral multiplets. Adding these up, we arrive at
$81$ twisted sector chiral multiplets.
For $\ZZ2\times\ZZ3$, with generators 
$\alpha=(\fr12,\fr12,0),\beta=(0,\fr13,\fr23)$, we have the following
states:
in the $\alpha$ twisted sector we have $16$ fixed tori; of these, four are 
singlets under the $\ZZ3$ group, the other $12$ sit in four triplets. The 
singlets contribute only half a torus cohomology invariant under $\beta$; 
for the triplets one can construct linear combinations of $0-$, $1-$ and 
$2-$forms that are invariant under $\beta$, giving a whole torus
cohomology 
per triplet. This sector therefore produces $12$ chiral multiplets.
The sector twisted by $\beta$ has $9$ fixed tori, 3 singlets and three 
doublets of $\ZZ2$, making up $9$ chiral multiplets, and the same for 
$\beta^2$. Finally, the sectors twisted by $\alpha\beta,\alpha\beta^2$ 
have $12$ fixed points; together they produce $12$ chiral multiplets. 
This adds up to $42$ chiral multiplets for the $\ZZ2\times\ZZ3$ model.

The calculation for the other models is similar, but somewhat more work. 
The results are 
\\

\noindent
\begin{tabular}{c|c}
model & number of chiral multiplets \\ \hline
$\ZZ3\times\ZZ3$ & $81$ \\
$\ZZ3\times\ZZ2$ & $42$ \\
$\ZZ2\times\ZZ6$ & $50$ \\
$\ZZ3\times\ZZ6$ & $71$ \\
$\ZZ6\times\ZZ6$ & $81$ \\
\end{tabular}
\\
 
\subsection{The massless open string spectrum}

We will now determine the massless states arising in the open string
sector. Different states occur depending on whether the string ends on
nine-branes ($99$-sector) or $5_i$-branes ($5_i5_i$-sector), which we will
call the unmixed sectors,  or whether the string has its two ends on
two different types of branes, the mixed sectors
($5_i9,5_i5_j$-sectors). In the unmixed sectors, the massless states in 
the bosonic (NS) sector are obtained by acting with
one oscillator on the ground state, which includes a Chan-Paton factor
$\lambda_{ij}$ labelling which branes the string ends on; the mixed sector
NS ground states are massless themselves. The form of
the Chan-Paton factors is then obtained by demanding the states to be
invariant under the orientifold group.   

We will only analyse the situation where there are no Wilson lines on the
branes, and with all five-branes of a given kind located on one fixed
point. This will give the maximum gauge symmetry. Furthermore we will so
obtain a T self dual configuration, so that nine- and five-branes will
produce the same gauge groups and matter content. All $\lambda$
will be $32$ by $32$ hermitian matrices.

The constraints on the $\lambda_{ij}$ are as follows. In the
$99$-sector we have four distinct sets of states, depending on whether the
oscillator has an index in spacetime (the gauge bosons) or in one of the
three internal tori (two scalars each).
On all these oscillators $\Omega$ acts as $-1$, the orbifold group element
$(\fr{n}{N},\fr{m}{M})$ distinguishes the different categories. Demanding
the total state to be invariant we have
\\

\noindent
\begin{tabular}{|l|l|l|}
\hline
state & $(\fr{n}{N},\fr{m}{M})$ & $\Omega$\\ \hline
$\psi^\mu|0,ij>\lambda_{ij}$  & $\lambda=
\gamma_{n,m,9}\lambda\gamma_{n,m,9}^{-1}$&
$\lambda=-\gamma_{\Omega,9}\lambda^T\gamma_{\Omega,9}^{-1}$\\ \hline
$\psi^{1\pm}|0,ij>\lambda_{ij}$  & $\lambda=
\expm{n}{N}\gamma_{n,m,9}\lambda\gamma_{n,m,9}^{-1}$&
$\lambda=-\gamma_{\Omega,9}\lambda^T\gamma_{\Omega,9}^{-1}$\\ \hline
$\psi^{2\pm}|0,ij>\lambda_{ij}$  & $\lambda=
\exmp{n}{N}\expm{m}{M}\gamma_{n,m,9}\lambda\gamma_{n,m,9}^{-1}$&
$\lambda=-\gamma_{\Omega,9}\lambda^T\gamma_{\Omega,9}^{-1}$\\ \hline
$\psi^{3\pm}|0,ij>\lambda_{ij}$  & $\lambda=
\exmp{m}{M}\gamma_{n,m,9}\lambda\gamma_{n,m,9}^{-1}$&
$\lambda=-\gamma_{\Omega,9}\lambda^T\gamma_{\Omega,9}^{-1}$\\ \hline

\end{tabular}
\\

\noindent The $55$-sectors give the same constraints, apart from an extra
minus in the action of $\Omega$ on the directions orthogonal to the
five-brane. In the mixed sectors, $95_i$ and $5_i5_j$, the NS ground state
is a massless spinor in the four directions with ND boundary conditions.
Considering for example the $5_19$ sector, these will be the $6789$
directions. There will be two states, labelled by their spin in the two
tori, which are equal due to GSO projection. There is no constraint on the
Chan-Paton matrix from the orientation
reversal: this relates the $5_19$ state to a $95_1$ state. We only have   
\\

\noindent
\begin{tabular}{|l|l|}
\hline
state & $(\fr{n}{N},\fr{m}{M})$\\ \hline
$|s_3=s_4,ij>\lambda_{ij}$ &
$\lambda=e^{2\pi i \frn s_3}\gamma_{n,m,5}\lambda\gamma_{n,m,9}^{-1}$\\
\hline
\end{tabular}
\\

\noindent and similar states for the other sectors.

Now we can use the representations of the $\gamma$'s we found to determine
the representations of these states. The results are collected in the
table.
\\

\noindent
\begin{tabular}{|l|lll|}
\hline
model &&& gauge group and matter\\ \hline
$\ZZ3\times\ZZ3$ & $99:$&$\psi^\mu:$&$ U(4)^3\times SO(8)$\\ &&
$\psi^1:$&$
(1,6,1,1),
(\ol{4},1,4,1),(1,\ol{4},1,8)$ \\
&& $\psi^2:$&$ (\ol{6},1,1,1),
(1,4,4,1),(4,1,1,8)$\\ 
&& $\psi^3: $&$(1,1,6,1),
(\ol{4},4,1,1),(1,1,\ol{4},8)$ \\\hline
$\ZZ3\times\ZZ6$ & $99\& 55:$& $\psi^\mu:$&$U(2)^6\times U(4)$\\
&&$\psi^1:$&$(2,\ol{2},1^5)$\footnotemark[1] 
$,(1^2,\ol{2},2,1^3),(1^4,\ol{2},2,1),$\\&&&$(1^4,2,1,\ol{4}),(1^5,\ol{2},4)$\\
&&$\psi^2:$&$(1,2,1^3,2,1),(2,1^3,2,1^2),(1^2,2,2,1^3),$\\
&&&$(1^3,\ol{2},\ol{2},1^2),(1^2,\ol{2},1^2,\ol{2},1),(\ol{2},\ol{2},1^5)$\\
&&$\psi^3:$&$(1,1_2,1^5)$\footnotemark[2]$,(1^3,1_{-2},1^3),
(\ol{2},1^4,\ol{2},1),$\\
&&&$(1^2,2,1,2,1^2),(1,\ol{2},1^4,\ol{4}),(1^3,2,1^2,4)$\\ 
&$59:$&& $(2,1^6;1,\ol{2},1^5),(1^3,2,1^3;1^2,\ol{2},1^4), 
(1^5,2,1;1^4,\ol{2},1^2)$\\ &&& $(1^4,2,1^2;1^6,\ol{4}), 
(1^6,4;1^5,\ol{2},1),$\\
&&& plus the same with the gauge groups reversed.\\ \hline
$\ZZ2\times\ZZ3$ & $99\&55:$& $\psi^\mu:$& $U(4)\times U(4)\times U(8)$\\
&&$\psi^1:$& $(4,\ol{4},1),(\ol{4},1,8),(1,4,\ol{8})$\\ 
&&$\psi^2:$& $(\ol{6},1,1),(1,6,1),(4,1,8),(1,\ol{4},\ol{8})$\\ 
&&$\psi^3:$& $(4,4,1),(\ol{4},\ol{4},1),(1,1,28),(1,1,\ol{28})$\\ 
&$59:$&& $(4,1^2;1,\ol{4},1),(1^2,8;\ol{4},1^2), (1,4,1;1^2,\ol{8})$\\
&&& plus the same with the gauge groups reversed.\\ \hline
$\ZZ2\times\ZZ6$&$99\&5_i5_i$ &$\psi^\mu:$& $U(4)\times USp(8)$\\
&&$\psi^1:$& $(16,1),(1,28)$\\
&&$\psi^2:$& $(\ol{6},1),(4,8)$\\
&&$\psi^3:$& $(6,1),(\ol{4},8)$\\
&$5_19\&5_25_3$ && $(4,1;\ol{4},1), (\ol{4},1;4,1), (1,8;1,8)$\\
&$5_29\&5_15_3$ && $(\ol{4},1;\ol{4},1), (4,1;1,8), (1,8;4,1)$\\
&$5_39\&5_15_2$ && $(4,1;4,1), (\ol{4},1;1,8), (1,8;\ol{4},1)$\\ \hline
$\ZZ6\times\ZZ6$& $99\&5_i5_i$& $\psi^\mu$&$U(2)^3\times USp(4)$\\
&&$\psi^1:$&$(2,\ol{2},1,1),(1,1,1_{-2},1),(1,1,2,4)$\\
&&$\psi^2:$&$(2,1,\ol{2},1),(1,1_{-2},1,1),(1,2,1,4)$\\
&&$\psi^3:$&$(1,\ol{2},\ol{2},1),(1_2,1^3),(\ol{2},1,1,4)$\\
&$5_19\&5_25_3$&&$(2,1^3;1,\ol{2},1^2),(1,\ol{2},1^2;2,1^3),(1,1,2,1;1^3,4),$\\
&&&$(1^2,\ol{2},1;1^2,\ol{2},1), (1^3,4;1,1,2,1)$\\
&$5_29\&5_15_3$&&$(1,\ol{2},1^2;1,\ol{2},1^2),(2,1^3;1^2,\ol{2},1),
(1^3,4;1,2,1^2),$\\
&&&$(1,1,\ol{2},1;2,1^3), (1,2,1^2;1^3,4)$\\
&$5_39\&5_15_2$&&$(2,1^3;2,1^3),(1,\ol{2},1^2;1^2,\ol{2},1),(\ol{2},1^3;1^3,4),$\\
&&&$(1^2,\ol{2},1;1,\ol{2},1^2), (1^3,4;\ol{2},1^3)$\\
\hline
\end{tabular}
\\

\footnotetext[1]{the notation $1^2$ is short for $1,1$, etc.}
\footnotetext[2]{the subscript indicates $U(1)$ charge; these are
antisymmetric tensors of
$U(2)$: 
singlets under $SU(2)$, but charge $2$ under
$U(1)$.}
\noindent This concludes the calculation of the spectra. 

\section{Discussion}

We have calculated the spectra of certain four-dimensional orientifold
models with $N=1$ supersymmetry. The spacetime symmetry divided
out was of the form $\ZZ{N}\times\ZZ{M}$, with the first factor acting on
the $4567$ directions, and the second on the $6789$ directions. We found
solutions to the tadpole equations in most cases, and used these to
compute the open string massless spectrum. We also computed the closed
string twisted sectors in these cases. Unexpectedly, two of the models
under consideration, $\ZZ2\times\ZZ4$ and $\ZZ4\times\ZZ4$, turned out not
to allow
a solution to the tadpole equations that satisfied the consistency
conditions. 

The orientifold models are expected to have dual heterotic orbifolds. This
follows from the ten-dimensional heterotic--type I duality \cite{PW}. In
fact, such
duals were constructed for the $N=M=3$ case in \cite{KS, AFIUV}. It would
be interesting to see whether a heterotic approach to the two inconsistent
models may shed some light on the reason for their inconsistency.

We have not exhausted all possibilities for orientifolds with
$\ZZ{N}\times\ZZ{M}$ orbifold group. It is for instance possible to
introduce discrete torsion, or to have models where the groups
act differently on the tori.     

\section*{Acknowledgements}

The author thanks Erik Verlinde for advice and discussions. This work is
supported by FOM.

\begin{appendix}

\setcounter{equation}{0}

\section{The tadpole calculation}
We briefly discuss the calculation of the RR-tadpoles due to the
orientifolds and branes. Demanding them to vanish gives equations for the
unitary matrices $\gamma$ representing the action of the orientifold group
on the Chan-Paton factors.  

The RR-charges can most easily be calculated by considering the exchange
of RR closed strings between the various orientifold planes and 
D-branes.
Schematically the amplitude for such a process is
$$
{\cal A}= \int dl \sum <boundary_1|e^{-2\pi l(p^2+m^2)}|boundary_2>,
$$
Here $l$ denotes the proper time along the tube, the sum is over all
RR-states propagating in the loop, and the boundaries are crosscaps for
orientifold planes, and normal boundaries for D-branes. In the limit
$l\to\infty$ the massless exchange diverges; the above
diagram is then proportional to $\int dl Q_1Q_2$, the RR-charges of the
boundaries. If we take both boundaries to be the combined state of all
orientifold planes and D-branes, this is proportional to the total charge
squared, which should vanish since the space transverse to the branes
and planes is compact.  

The various closed string exchange diagrams can be reinterpreted as open
string loops. Two crosscaps give a Klein Bottle (KB) diagram, one crosscap
and
a boundary a M\"obius strip (MS), and two boundaries a cylinder (C). The
RR-channel states then correspond to the open string traces with $(-1)^F$
for the cylinder, the R-sector for the MS and R with $(-1)^F$ and NS 
for the
KB. (Note, however, that due to the zero-modes the traces with 
$(-1)^F$ in the R-sector vanish). A 
consistency check is that if we add up all diagrams (per twisted
sector) we should get a square, since then we are computing the total
(orientifold and brane) charge squared.

We will use the following notation. The orientifold we are interested in
has orbifold group $\ZZ{N}\times\ZZ{M}$, which acts on the three complex
coordinates of the six-torus as 
\be
(z_1,z_2,z_3)\to (e^{\frac{2\pi i n}{N}}z_1, e^{-\frac{2\pi
i n}{N}}z_2,z_3),\mbox{ resp. } (z_1,z_2,z_3)\to (z_1,e^{\frac{2\pi i
m}{M}}z_2, e^{-\frac{2\pi i
m}{M}}z_3).
\ee
The branes involved are nine-branes, and possibly five-branes stretched
along four-dimensional spacetime plus one of the three internal tori; they
are referred to as $5_1,5_2$ and $5_3$. The $\gamma$'s carry indices $n,m$
and possibly $\Omega$ to denote the group element, and a $9$ or $5_i$ to
indicate in which sector they act.

One has to compute the various loop amplitudes, which are traces with 
insertion of the different orbifold group elements, and a factor $\Omega$ 
for MS and KB. The traces include a trace over the Chan-Paton indices 
when the diagram has a boundary on the different D-branes (i.e. for the C 
and MS diagrams). The results can be conveniently expressed in terms of 
Jacobi's $\vartheta$-functions (for their definition, see e.g. \cite{GJ}). 
From supersymmetry one has that the sum of NS and R-sectors, with and 
without $(-1)^F$, vanishes for each diagram. In the calculation, this is 
a consequence of the various identities for sums of products of 
$\vartheta$-functions. 

From the amplitude, one then extracts the divergent behaviour in the long 
tube limit. This limit coincides with the limit of the modular parameter 
of the open string diagram going to zero. The asymptotics can be 
calculated using the transformation properties of the theta-functions 
under modular transformations \cite{GJ}. 

As an example we give the amplitude for the cylinder diagram. Here 
different cases can be distinguished: either boundary of the cylinder can 
lie on a nine-brane or on a five-brane. To be specific let us take both 
boundaries on a nine-brane. In the sector where we include the group 
element labelled by $(n,m)$ (so that this twisted RR field strength 
propagates in the loop in the closed string tree channel perspective) we 
should calculate the trace
\be
\Tr \left( \frac{1}{2NM}\alpha^n\beta^m\frac{1+(-1)^F}{2}e^{-2\pi t L_0} 
\right) 
\ee
The trace is over both NS and R sector states, the latter contributing
with a minus sign. $\alpha$ and $\beta$ are the generators of the 
orbifold group; the factor in front is the division by the order of the 
orientifold group.  

The total amplitude in this case is given by the expression
\bean
&&\A^{\mbox{C}}_{99}(n,m) = \frac{v_4}{16NM}(\Tr{\gamma_{n,m,9}})^2 
\left(\frac{1}{8\sin\frac{\pi n}{N}\sin\frac{\pi
m}{M}\sin\pi (\frac{m}{M}-\frac{n}{N})}\right)^2\\ &&\int
\frac{dt}{t^3} 
\sin\pi (\frn-\frm) 
f_1(t)^{-3}\th{1}{\frn}^{-1}\th{1}{\frm}^{-1}\th{1}{\frn-\frm}^{-1}  
\times \\ &&\Bigg(
\th{3}{0}\th{3}{\frn}\th{3}{\frn-\frm}\th{3}{\frm}
- \th{4}{0}\th{4}{\frn}\th{4}{\frn-\frm}\th{4}{\frm}- \\
&& \th{2}{0}\th{2}{\frn}\th{2}{\frn-\frm}\th{2}{\frm} \Bigg) .
\eean
The factor $v_4$ in front is the regularised dimension of the non-compact 
directions, in string units; it arises from the integration over the 
non-compact momenta, which is part of the trace. The gamma-matrices 
reflect the action of the group element on the nine-brane Chan-Paton 
indices. The subsequent goniometric factor comes from the action of the 
group element on the compact space \cite{GJ}, and is equal to one over the 
number of fixed points. Under the integral sign we find the oscillator 
contributions. The inverse theta-functions represent the bosonic 
contributions, which appear in the denominator. Their definitions include 
goniometric functions that have to be cancelled, hence the three sines. 
The arguments of the theta-functions show the action of the group element on 
the associated complex coordinates. The function 
$f_1$ is 
associated to oscillators in the uncompact directions; its definition can 
also be found in \cite{GJ}. The expression in brackets represents the 
fermionic oscillators, appearing in the numerator. The first term comes 
from the NS-sector, the second one from the NS-sector with the 
inclusion of $(-1)^F$ in the trace, while the third one is the R-sector 
(with an additional minus sign because of the fermion loop). The total 
expression vanishes; the contribution of the RR exchange only is given by 
the second term. 

The massless RR exchange is obtained by taking the limit of infinite tree 
channel parameter $l$, which in the case of the cylinder is related to 
$t$ as $t=\fr1{2l}$. 
This will be proportional to the nine-brane charge w.r.t. the twisted 
RR-potential squared. The limiting behaviour can be obtained using the 
transformation properties of the Jacobi functions under modular 
transformations. 

Similar computations were carried out for the other sectors ($5_i9$, 
$5_i5_j$), as well as for the MS and KB amplitudes. For the latter two, 
the element $\Omega$ is to be added in the trace; the diagrams involving 
the group element $\alpha^n\beta^m$ are associated to exchange of 
the $2n,2m$-twisted RR-potential, between orientifold-plane and D-brane 
(in the case of the MS), or between two orientifold-planes (KB). Hence,
different diagrams contribute to the various channels depending on whether
$N$, $M$ are odd or even. These diagrams give contributions involving 
the charge of the orientifold-planes. 

In the end, the total charge of planes and branes with respect to all 
different twisted RR-potentials should vanish. This gives a constraint on 
the traces of the different $\gamma$-matrices. Let us start with the case 
where $N,M$ are both even. In the odd-twisted sectors, i.e. $n$ or $m$ is
odd, we only get a contribution from the cylinders, proportional to:
\bea
\label{eq:ee1}
&&\frac{1}{\sin\pi\frn \sin\pi \frm\sin \pi|\frn-\frm|} \times \\ \nn 
&&\Big[
\Tr\gamma_{n,m,9} 
 + 4\sin\pi \frm\sin \pi(\frn-\frm)\Tr \gamma_{n,m,5_1} \\ \nn && -
4\sin\pi \frn\sin \pi\frm \Tr \gamma_{n,m,5_2} - 4\sin\pi
\frn\sin
\pi(\frn-\frm)\Tr
\gamma_{n,m,5_3}\Big]^2
\eea

In the case where both $n$ and $m$ are even, there are also contributions
of the $MS$ and twisted and untwisted $KB$ amplitudes. We find a
contribution proportional to
\bea
\label{eq:ee2}
&& \frac{1}{\sin 2\pi \frn\sin 2\pi \frm\sin 2\pi
(\frn-\frm)}\times \\ \nn
&&\Big[\Tr\gamma_{2n,2m,9} + 4 \sin 2\pi \frm\sin
2\pi
(\frn-\frm)\Tr
\gamma_{2n,2m,5_1}\\ \nn
&&- 4 \sin 2\pi \frn\sin 2\pi \frm\Tr
\gamma_{2n,2m,5_2} - 4 \sin 2\pi \frn\sin 2\pi (\frn-\frm)\Tr
\gamma_{2n,2m,5_3} \\ \nn
&&  - 32(\cos 2\pi \frn \cos^2 \pi \frm - \sin^2 \pi
\frm + \half \sin 2\pi \frn\sin 2\pi \frm)\Big]^2.
\eea
For this we used that 
$$
\Tr(\gamma_{m,n,\Omega}\gamma_{m,n,\Omega}^{-T})=\pm \Tr\gamma_{2m,2n}.
$$ 
When $n=0$ we have to include momentum and winding states along the
$4-5$-torus; the result then is
\be
\label{eq:ee3}
\frac{1}{\sin^2 \pi \frm}\Big[ \Tr
\gamma_{0,m,9}-4\sin^2 \pi \frm\Tr \gamma_{0,m,5_1}\Big]^2 
\ee
and
\be
\label{eq:ee4}
\Big[\Tr\gamma_{0,m,5_2} - \Tr
\gamma_{0,m,5_3}\Big]^2.
\ee
for odd $m$, while for even $m$ we find
\be
\label{eq:ee5}
\frac{1}{\sin^2 2\pi \frm}
\Big[ \Tr\gamma_{0,2m,9} - 4\sin^2 2\pi \frm \Tr \gamma_{0,2m,5_1}-32\cos 
2\pi \frm \Big]^2,
\ee
and
\be
\label{eq:ee6}
\Big[\Tr\gamma_{0,2m,5_2} - \Tr
\gamma_{0,2m,5_3}\Big]^2.
\ee
Similar expressions are found for the $m=0$ and $m=n$ sectors.
Finally, the expression for $m=n=0$ guarantees that there are $32$ branes
of each kind. 

The result when $M,N$ are both odd (so we have no five-branes, just
nine-branes) is

\be
\label{eq:oo1}
 \frac{1}{\sin 2\pi \frn\sin 2\pi \frm\sin 2\pi
(\frn-\frm)}\Big[\Tr\gamma_{2n,2m,9} 
  - 32\cos \pi \frn \cos \pi \frm \cos \pi (\frn-\frm) \Big]^2.
\ee
Again for $n=0$ we obtain
\be
\label{eq:oo2}
\frac{1}{\sin^2 2\pi \frm}
\Big[ \Tr\gamma_{0,2m,9} - 4\sin^2 2\pi \frm \Tr
\gamma_{0,2m,5_1}-32\cos^2 \pi \frm \Big]^2,
\ee
while the $n=m=0$ result tells us that there are $32$ nine-branes and no
five-branes.

Finally, when $N$ is odd, $M$ even, we have, for odd $m$
\be
\label{eq:eo1}
 \frac{1}{\sin\pi \frn\sin\pi \frm\sin \pi|\frn-\frm|}
\Big[
\Tr\gamma_{n,m,9} + 4\sin\pi \frm\sin \pi(\frn-\frm)\Tr
\gamma_{n,m,5_1}\Big]^2,
\ee
and for even $m$
\bea
\label{eq:eo2}
&& \frac{1}{\sin 2\pi \frn\sin 2\pi \frm\sin 2\pi
(\frn-\frm)}\times \\ \nn 
&&\Big[\Tr\gamma_{2n,2m,9} + 4 \sin 2\pi \frm\sin
2\pi
(\frn-\frm)\Tr
\gamma_{2n,2m,5_1}\\ \nn   
&&  - 32(\cos \pi \frn \cos\pi \frm\cos\pi (\frn-\frm)+\cos\pi \frn\sin
\pi
\frm\sin\pi (\frn-\frm))\Big]^2.
\eea
If $n=0$ the answer is that of the even $N,M$ situation, eqs.
(\ref{eq:ee3}--\ref{eq:ee6}), while for $m=0$
we find
\be
\label{eq:eo3}
\frac{1}{\sin^2 2\pi \frn}
\Big[ \Tr\gamma_{2n,0,9} -32\cos^2 \pi \frn \Big]^2 
\ee
and
\be
\label{eq:eo4}
\Big[\Tr\gamma_{2n,0,5_1} - 8 \Big]^2.
\ee
Finally we again should have $32$ nine- and $5_1$-branes for the untwisted
channel cancellation.

Requiring all these expressions to vanish, one obtains the tadpole
equations. The solutions to these equations are presented in section
\ref{sec:solns}.

\end{appendix}


\begin{thebibliography}{99}
\bibitem{DHVW}
L. Dixon, J. Harvey, C. Vafa, E. Witten, ``Strings on Orbifolds 1 \& 2'',
\NPB{261}{85}{678}  and \NPB{274}{86}{285}.
\bibitem{Ibetal}
L.E. Ibanez, J. Mas, H-P. Nilles, F. Quevedo, ``Heterotic Strings in
Symmetric and Asymmetric Orbifold Backgrounds'', \NPB{301}{88}{157};\\
A. Font, L.E.Ibanez, F.Quevedo, A.Sierra, ``The Construction of
'Realistic' Four-Dimensional Strings through Orbifolds'',
\NPB{331}{90}{421};\\
A. Font, L.E. Ibanez, F. Quevedo, ``$\ZZ{N}\times\ZZ{M}$ Orbifolds and
Discrete Torsion'' \PLB{217}{89}{272}. 
\bibitem{Dbranes}
S. Chaudhuri, J. Polchinski, C.V. Johnson, ``Notes on D-Branes'',
hep-th/9602052.
\bibitem{ori}
          See, {\it e.g.}\/:\\
       J. Dai, R.G. Leigh, and J. Polchinski, ``New Connections between
String Theories'', 
      {\it Mod.\ Phys.\ Lett.}\/ {\bf A4} (1989) 2073;
   A. Sagnotti, ``Open Strings and their Symmetry Groups'', in Proceedings
of {\it Cargese 1987: Non-Perturbative
   Quantum Field Theory}\/, eds.\ G. Mack \etal\  (Plenum, 1988), p.\
521;\\
   P. Ho\v{r}ava, ``Strings on World Sheet Orbifolds'', 
\NPB{327}{89}{461}; \PLB{231}{89}{251};\\
   G. Pradisi and A. Sagnotti, ``Open String Orbifolds'',  
\PLB{216}{89}{59};\\
   M. Bianchi and A. Sagnotti, ``On the Systematics of Open String
Theories'', 
       \PLB{247}{90}{517}; ``Twist Symmetry and Open
String Wilson Lines'', \NPB{361}{91}{519};\\
    A. Sagnotti, ``Some Properties of Open String
Theories'', hep-th/9509080;\\
\bibitem{GP}
E.G. Gimon, J. Polchinski, ``Consistency Conditions for Orientifolds and D
Manifolds'', \PRD{54}{96}{1667}, hep-th/9601038.
\bibitem{PW}
J. Polchinski, E. Witten, ``Evidence for Heterotic -- Type I String
Duality'', \NPB{460}{96}{525}, hep-th/9510169. 
\bibitem{GJ}
E.G. Gimon, C.V. Johnson, ``K3 Orientifolds'',
\NPB{477}{96}{715}, hep-th/9604129.
\bibitem{DP}
A. Dabholkar, J. Park, ``Strings on Orientifolds'',
\NPB{477}{96}{701}, hep-th/9604178.
\bibitem{BL}
M. Berkooz, R.G. Leigh, ``A $D$=$4$ $N$=$1$ Orbifold of Type I Strings'',
\NPB{483}{97}{187}, hep-th/9605049.
\bibitem{VW}
C. Vafa, E. Witten, ``On Orbifolds with Discrete Torsion'',
\JGP{15}{95}{189}, hep-th/9409188.
\bibitem{K}
Z. Kakushadze, G. Shiu, ``A Chiral $N$=$1$ Type I Vacuum in Four
Dimensions
and its Heterotic Dual'', \PRD{56}{97}{3686}, hep-th/9705163;\\
Z. Kakushadze, ``Aspects of $N$=$1$ Type I Heterotic Duality in Four
Dimensions'', hep-th/9704059;\\
C. Angelantonj, M. Bianchi, G. Pradisi, A. Sagnotti, Ya. S. Stanev,
``Chiral Asymmetry in Four-dimensional Open String Vacua'', 
\PLB{385}{96}{96}, hep-th/9606169.
\bibitem{KS}
Z. Kakushadze, G. Shiu, ``$4$-$D$ Chiral $N$=$1$ Type I Vacua with and
without
D5-Branes'', hep-th/9706051.
\bibitem{AFIUV}
G. Aldazabal, A. Font, L.E. Ibanez, A.M. Uranga, G. Violero,
``Non-Perturbative Heterotic $D$=$6$, $D$=$4$, $N$=$1$ Orbifold Vacua'',
hep-th/9706158.

\end{thebibliography}
\end{document}